\documentstyle[12pt]{article}

         \def\be{\begin{equation}}
         \def\bea{\begin{eqnarray}}

         \def\ee{\end{equation}}
         \def\eea{\end{eqnarray}}
         \def\R{\rm {I\kern-.200em R}}
         \def\C{\rm {I\kern-.520em C}}
\begin{document}
\begin{center} {\large \bf Large Extra Dimensions and Noncommutative Geometry\\
in String Theory}\\

\vskip 1cm
Farhad Ardalan \footnote{ardalan@theory.ipm.ac.ir}\\
School of Physics, IPM\\   P.O. Box 19395-5531, Tehran-Iran\\
and\\
Department of Physics, Sharif University\\
P.O. Box 11365-9161, Tehran-Iran
{}\\
\vskip 1cm
\end{center}

\vskip 1cm

\begin{abstract}
The consequences of noncommutativity of space coordinates of string 
theory in the  proposed large
extra dimension solution to the hierarchy problem are explored; in particular 
the large
dimension stabilization and the  graviton  reabsorption in the brane are considered.
\end{abstract}
\newpage

In ordinary quantum mechanics, space coordinates of a point or 
separate points commute and can therefore be simultaneously measured. However,
it has long been suspected that in the realm where quantum effects of gravity become 
significant, space coordinates may not commute; and it has been considered 
ever  since the inception of quantum mechanics.

It is only recently that a thorough mathematical theory of noncommutative 
geometry has been developed [1], accommodating the simple notion of 
noncommuting coordinates, the application of which to various aspects has been 
extensive [2]; the most significant application being to quantum gravity
through string field theory [3], and branes in string theory [4].
The discovery of the noncommutative geometry of the branes in string
theory has spurred a flurry of research activity in noncommutativity in string
theory (See reference 4 for a detailed review).

The appearance of noncommutativity in the presence of branes in string theory 
is closely tied to the non vanishing expectation value of the NS-NS axionic two 
form
field B with components along. When the field B has non vanishing components on the 
D-brane, the coordinates of the open strings attached to them and the coordinates 
along the D- Branes fail to commute. In certain limits of the theory where gravity
is decoupled,  this noncommutativity results in appearance of a nontrivial gauge
theory whose properties have been under extensive study.

In a separate development, a novel solution to the problem of hierarchies
between the Planck scale of $10^ {19} GeV$ and the electroweak  scale 
of about 1 TeV, has been recently proposed via the existence of submillimeter
extra space dimensions beyond the three physical space dimension of 
our universe [5].
In this scenario the extra dimensions acquire submillimeter lengths and have
dramatic  consequences in both classical gravity experiments and in 
future collider experiments. Studies on the effects of these extra dimensions 
on  astrophysical and cosmological observables, in addition to the present data 
on classical gravity and collider experiments, have yielded bounds on the 
scale
and the number of extra dimensions.

An outstanding problem in the proposed large extra dimension solution of the 
hierarchy problem is the dynamical stabilization of the extra dimensions, for 
which certain possibilities have been proposed. Another problem that arises 
in the models of extra dimension is the question of the graviton and other
bulk fields reabsorption on the three brane forming our world.In this letter 
consequences of the noncommutativity of space coordinates 
of the branes on the scenario of large extra dimension is noted and its 
bearing on the above two problems are discussed.

To begin with, following reference [4], note that noncommutativity in
D-branes appear as soon as a non zero constant Kalb-Ramond B field
is turned on: The world sheet action  of the open strings in the presence
of the B field

\be S= {\frac {1} {4 \pi \alpha^\prime}} \int _\Sigma (g_{ij} \partial _a X^i \partial ^a X^j
-2 \pi \alpha ^\prime B_{ij} \epsilon ^{ab} \partial _u X^i \partial _b X^j)+ \int 
_{\partial \Sigma} A_i \partial X^i\ee
with $\Sigma$ the string world sheet and $g_{ij}$, the metric of the target space, and
A the gauge field at the end of the open string, leads to the mixed boundary
condition along the D-brane:

\be g_{ij} \partial _\bot X^j + 2 \pi \alpha ^\prime B_{ij} \partial_t X^j=0 \ee
The action (1) is invariant under two independent gauge transformations;

\be \delta A_i = \partial _i \lambda,\ee

and

\be \delta A_i= -\Lambda _i \ee

\begin{center} $\delta B = d \Lambda $ \end{center}

Quantizing the string with this boundary condition is in consistent
with the usual canonical commutation relations of the open string [6].
A consistent Dirac quantization forces some coordinates of the
open strings to be noncommutative whose location depend on the fixing
of the gauge invariance (4).In the particular gauge of dA=0, the end 
points of the open strings,
i.e. the coordinates of the brane along the brane, become noncommutative;

\be [X_i, X_j]=\frac {i} {2} \Theta _{ij}\ee

where

\be \Theta _{ij}=2 \pi \alpha^ \prime (\frac {1} {g+2\pi \alpha ^\prime B})^{ij}_{anti} \ee
In the simplest case of two dimensions

\be \Theta _{12}=  \frac {(2 \pi \alpha \prime B)^2} {det g+ (2\pi \alpha \prime B)^2} 
\frac {1} {B}\ee 
where $B = B_{12}$. We will soon explore the consequences of this 
noncommutativity of the two space coordinates $X_1$ and $X_2$ along the 
brane in the context of the large extra dimension proposal.

In the string theoretical version of the large extra dimension proposal 
our three dimensional 
space is a sub manifold of a brane on which, naturally, the gauge fields reside 
and the other fields of the standard model of particle physics
live. There are graviton and other fields which propagate in the full
nine dimensional bulk of the space of string theory (or the ten dimensional 
space of M-theory). Of these bulk fields the Kalb-Ramond B field plays an
important role in our discussions. According to the analysis of ref. [5]
the gravitational strength in the four dimensional world is considerably reduced 
compared to that of the bulk, provided the size R of the extra compact 
dimensions is sufficiently larger than the string scale, i.e.,

\be M^2 _{pl} = M^2 _* (M_* R)^ n \ee

Here $M_*$ is the string mass scale  and $M_{pl}$ is the ordinary four dimensional
Planck mass; n is the number of compact dimensions. Correspondingly the coupling 
constant of various interactions are reduced by the volume effect, $(M_* R)^n$,
compared to the bulk.

To explore the effect of the above mentioned noncommutativity  (7) on the 
large extra dimension proposal, let us consider the case of a 5-brane
wrapped on a two torus with coordinates $X_1$ and  $X_2$. In order to conform  
with the large extra dimension scenario, the coordinate $X_1$ and $X_2$
should be of the weak (string) scale so that the gauge fields and other 
standard model fields are initially confined to the narrow 3-brane with
the usual $M_*^{-1}$ thickness [7]. The remaining 
three dimensional sub manifold of the 5-brane constitutes our world. Some or all 
of the other 
dimensions of the space are then compactified at the large scale.This is 
a simple situation for which the
phenomena of space coordinate noncommutativity shows up, if we allow the 
B field to have a non vanishing expectation value along the $X_1$ and $X_2$ 
directions.

As a result of noncommutativity of $X_1$ and $X_2$, eq. (5), these coordinates can 
not be simultaneously set to a certain value, e.g. zero, and the 3-brane 
of our world develops a "thickness" along the extra dimension $X_1$ and $X_2$.
As long as this thickness $\Delta$, is comparable to the string scale 
(weak scale) $M_*^{-1}$, no new phenomenon is expected. However, under certain 
conditions the size of the thickness $\Delta$ can be much larger than  $M_*^{-1}$
and it will have drastic consequences on the phenomenological implications of
the large extra dimension scenario.

In analogy to the phase space noncommutativity of ordinary quantum 
mechanics, where $[X,p]= i \hbar$, the size of the thickness $\Delta$ will be
proportional to $\theta _{12}$. And from eq. (7), it is easily seen that $\theta$
can have as large a value as 

\be \Delta _{max} \sim \theta _{max} \sim \frac{\alpha ^\prime B}{det g} \ee
In the usual large extra dimension scenario, B is simply the bulk decay 
constant of axion and must be of the order $M_*$[5] and therefore

\be \Delta _{max} \sim \frac {M_*^{-1}} {det g};\ee
then for small volumes, $det g$, of the $X_1$, $X_2$ space, a large thickness 
develops, even though we started from weak scale sizes for $X_1$ and $X_2$.

When the B field has non zero component both along the 3-brane and connecting 
the 3-brane and the other directions of the 5-brane, there will be  large 
corners of the moduli space for which $\theta _{12}$ and consequently 
$\Delta$, will be large.

The most immediate result of a large $\Delta$ will be an enhancement of the 
cross section for 
reabsorption of the bulk fields such as gravitons, by the 3-brane. In the 
original discussion of the cosmological consequences of the large extra dimension
theory [5], there is a bound on the reheating temperature $T_*$, due to 
over closure of the universe because of graviton emission into the bulk. The bulk
gravitons are partially reabsorbed by the three brane and decay into two photons.
The observed present decay photon flux then puts a bound on $T_*$ which falls
far short of the temperature for the critical mass and can not account for the 
dark matter. The solution envisaged  in Ref[5] is to allow for another, preferably 
of a larger dimension brane on which gravitons may decay, which in turn will 
reduce the 
branching ratio for the graviton decay on the 3-brane.

Now when the noncommutativity  of the coordinates $X_1$ and $X_2$ force 
the 3-brane to be thick with a large volume, the gravitons will have a much 
shorter lifetime and the reheating temperature will accordingly be lowered 
and the possibility of accounting for dark matter will further rescind.

The noncommutativity of the coordinates orthogonal to the 3-brane of our world will 
probably be most important for the stabilization of these coordinates. The point 
is that during the cosmological evolution of the dimensions, there will be 
a minimum size, due to the noncommutativity of the coordinates, in which  the
effective potential for the extra dimensions will have a minimum.
There are certain models for the stabilization of the dimensions which are 
relevant in this context.  One model [8], tailored for this purpose
in five spacetime dimensions, achieves stabilization of the one extra dimension
through use of a bulk scalar field $\varphi (x,y)$. In this model the scalar
bulk field is added with a certain potential of the generic form

\be V(\varphi)= \int \sqrt {detg} (\varphi^2 - \upsilon^2)^2 \ee
and allowed to develop a non vanishing vacuum expectation value which 
depends on the extra dimension. The integration over the extra dimension
will result in an "effective" potential for the parameters of V and g, which 
include
the size of the extra dimension.
This procedure can be generalized for the case of more than one extra 
dimension, where now the two "parameters" of scale in the
two extra dimensions do not commute. The minimum of the effective potential is 
then expected to be at a scale of $\Delta$.

In whatever manner the effective potential for the size of the extra dimensions
is generated, the noncommutativity of these coordinates presents a natural
reason for the existence of a minimum size of the order of $\Delta$ for the extra
dimension. For example in the asymmetric inflation model of ref[9], where a
potential $V (\varphi)$ is assumed for the inflaton field $\varphi$, which 
is 
in turn identified with the extra dimension size the radion, again the 
noncommutativity of the coordinates sets a lower bound on $\varphi$. moreover
the moduli problem which appear in this model, i.e. the large energy of the
radion field in the bulk, is automatically less severe as the radion decay
to the brane fields is enhanced due to the increased thickness of the branes.

We hope to come back to a detailed study of these issues later.

{\bf Acknowledgement}

The author would like to thank H. Arfaei and Y. Farzan for discussions.
\pagebreak

\end{document}